\documentclass{article}

\usepackage{microtype}
\usepackage{graphicx}
\usepackage{booktabs} % for professional tables
\usepackage[skip=0cm,list=true,labelfont=it]{subcaption}

\usepackage{hyperref}

\usepackage[accepted]{icml2018}

\icmltitlerunning{A Case Study of Deep-Learned Activations via Hand-Crafted Audio Features}

\begin{document}

\twocolumn[
\icmltitle{A Case Study of Deep-Learned Activations \\ via Hand-Crafted Audio Features}



\icmlsetsymbol{equal}{*}

\begin{icmlauthorlist}
\icmlauthor{Olga Slizovskaia}{upf}
\icmlauthor{Emilia G{\'o}mez}{upf,erc}
\icmlauthor{Gloria Haro}{upf}
\end{icmlauthorlist}

\icmlaffiliation{upf}{Department of Information and Communication Technologies, Pompeu Fabra University, Barcelona, Spain}
\icmlaffiliation{erc}{European Commission - Joint Research Centre, Seville, Spain}

\icmlcorrespondingauthor{Olga Slizovskaia}{olga.slizovskaia@upf.edu}

\icmlkeywords{Machine Learning, Music, Convolutional Neural Networks, Explainability}

\vskip 0.3in
]



\printAffiliationsAndNotice{}  % leave blank if no need to mention equal contribution

\begin{abstract}
This work presents a method for analysis of the activations of audio convolutional neural networks by use of hand-crafted audio features. We analyse activations from three CNN architectures trained on different datasets and compare shallow-level activation maps with harmonic-percussive source separation and chromagrams, and deep-level activations with loudness and onset rate. 

\end{abstract}

\section{Introduction}
\label{introduction}

In this paper, we focus on {\it feature analysis} in the music domain. Our goal is to find similar patterns between the features (activations and activation maps) learned by a network and hand-crafted audio features, which are well understood in the literature. 
For that purpose, we analyse features from a dataset of user-generated recordings of different musical instrument performances. We address musical instrument recognition  as it is a well-defined task and it can be objectively evaluated. 

For feature attribution understanding, there are two major directions: (1) perturbation based algorithms, such as LIME \cite{Ribeiro2016}, Axiomatic Attribution \cite{Sundararajan} or Saliency Analysis \cite{Montavon}, and (2) gradient-based algorithms such as Guided Backpropagation \cite{simonyan2013deep,Montavon}, Class-Activation Mapping (CAM) \cite{zhou2016learning}, and Network Dissection \cite{Bau}. 
In music domain, SoundLIME \cite{Mishra} algorithm has been adapted from the original LIME. However, in most cases, the above techniques can be limitedly applied to spectrograms because, unlike a typical image, two dimensions of a spectrogram represent different qualities namely time and frequency.

Therefore, manual feature exploration remains popular. One could create a playlist which corresponds to a particular neuron, and make a decision of this neuron 'specialization' by listening to the playlist. This approach was proposed by \cite{sanderspotify} and it provides valuable insights. However, it is not scalable because it requires an expert to listen to the playlist and guess the rationale behind. 

Also, we can take advantage of a number of well-established mid-level audio features that have been proposed and studied in the MIR literature \cite{midfeatures}. We know that CNNs in computer vision learn boundaries in the first layer and more complex concepts in subsequent layers. We hypothesize that audio-based CNNs can occasionally learn some of the hand-crafted features in a similar manner. We try to identify those features in pre-trained neural networks.   

\section{Methodology}

\textbf{Hand-crafted audio features.} We focus our study in a compact set of mid-level features related to different musical facets: onset rate, loudness and Harmonic Pitch Class Profile (HPCP) computed by \texttt{Essentia} \cite{essentia}, and Harmonic/Percussive Sound Separation (HPSS) computed by \texttt{librosa} \cite{librosa}.

\textbf{Network Architectures.} We explore three state-of-the-art VGG-style architectures: CNN AudioTagger (CNN-AT) \cite{choi2016automatic}, VGGish \cite{vggish2017}, and Musically Motivated CNN (MM-CNN) \cite{PonsTimbre}.
All three receive mel-spectrum as the input, consist of blocks of convolutional and max-pooling layers, and dense layers. 

The differences between architectures and their initializations include filters' shape (squared filters in CNN-AT and VGGish, and rectangular filters in MM-CNN), activation function and pre-training settings.  We trained CNN-AT and MM-CNN on a subset of FCVID \cite{FCVID} dataset. VGGish is initialized with weights provided by the authors. This network has been trained on a large-scale AudioSet dataset \cite{audioset} and potentially have stronger discriminative ability.

{\bf Similarity measures: individual activations}. For high-level embeddings of a network, we consider each activation as an individual feature and compare them with onset rate and mean loudness.
We consider two similarity metrics: (1) Pearson Correlation Coefficient and (2) Euclidean distance over the normalized vectors.

{\bf Similarity measures: activation maps}. Activations of convolutional layers have a form of a matrix. They are slightly offset from the original input spectrum due to the padding, and proportionally scaled to the input because of max pooling. To some extent, we can think of them as pseudo-spectrograms or as filtered and aggregated spectrograms. In order to compare those activations with HPSS or HPCP, we need a method for fuzzy matrix comparison which is scale- and shift-invariant. We propose a visual-inspired similarity metric based on Scale-Invariant Feature Transform (SIFT) \cite{LoweSIFT} descriptors. SIFT descriptors are among the most recognized features in computer vision and a reasonable choice for similarity measurement~\cite{hua2012similarity}. 

\begin{figure}
 \begin{subfigure}{\columnwidth}
    \includegraphics[width=\textwidth]{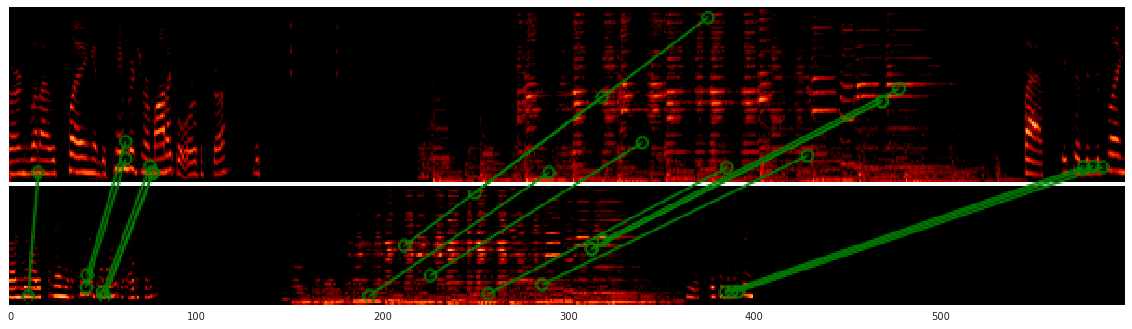}
    \caption{Top: original log-mel-spectrum. Bottom: logharmonic component of HPSS, scaled. SIFT matches are connected.}
 \label{fig:logharmonic_sift_example_scaled}
 \end{subfigure}%
 \\
 \begin{subfigure}{\columnwidth}
    \includegraphics[width=\textwidth]{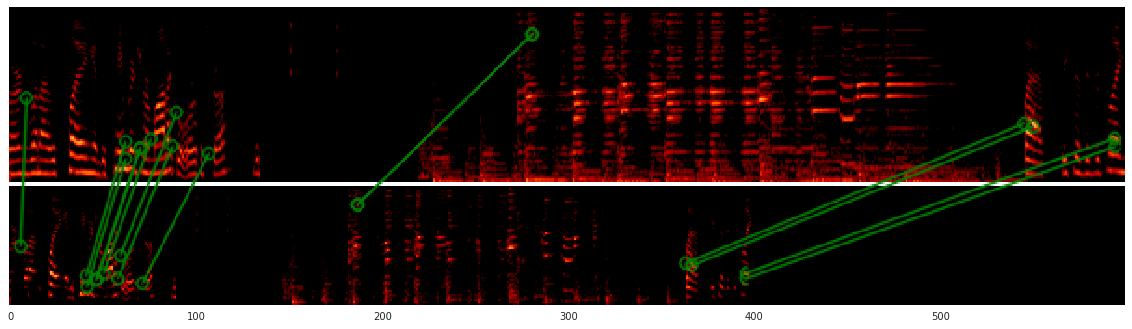}
    \caption{Top: original log-mel-spectrum. Bottom: logpercussive component of HPSS, scaled. SIFT matches are connected.}
 \label{fig:logpercussive_sift_example_scaled}
 \end{subfigure}%
     \caption{An example of SIFT matching for scaled harmonic (\ref{fig:logharmonic_sift_example_scaled}) and percussive (\ref{fig:logpercussive_sift_example_scaled}) parts of HPSS and shifted spectrum.}
    \label{fig:sift_example_scaled}
 \end{figure}

To compute similarity between a feature map and an activation map we compute SIFT descriptors and matches between descriptors.  
An example of matching is shown in Figure~\ref{fig:sift_example_scaled}. Each match is characterized by the matched descriptor indexes and a matching distance.

\section{Experiments and Results}

\begin{figure}
\centering
\begin{subfigure}{0.5\columnwidth}
\centering
	\captionsetup{justification=centering}
    \includegraphics[width=\textwidth]{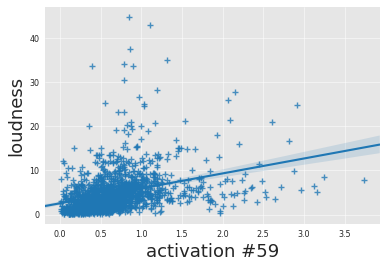}
    \caption{\small Loudness/Activation~\#59.}
    \label{fig:vggish_embedding_loudness_pcc}
\end{subfigure}%
\begin{subfigure}{0.5\columnwidth}
\centering
	\captionsetup{justification=centering}
    \includegraphics[width=\textwidth]{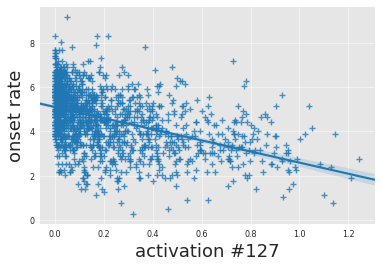}
    \caption{\small Onset rate/Activation~\#127.}
    \label{fig:vggish_embedding_onsetrate_pcc}
\end{subfigure}

\centering
\begin{subfigure}{0.5\columnwidth}
\centering
	\captionsetup{justification=centering}
    \includegraphics[width=\textwidth]{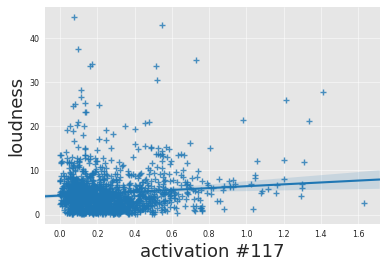}
    \caption{\small Loudness/Activation~\#117.}
    \label{fig:vggish_embedding_loudness_l2}
\end{subfigure}%
\begin{subfigure}{0.5\columnwidth}
\centering
	\captionsetup{justification=centering}
    \includegraphics[width=\textwidth]{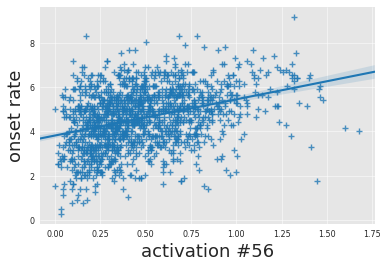}
    \caption{\small Onset rate/Activation~\#56.}
    \label{fig:vggish_embedding_onsetrate_l2}
\end{subfigure}
\caption{An example of correspondences between VGGish embeddings and mid-level audio features: \ref{fig:vggish_embedding_loudness_pcc} and \ref{fig:vggish_embedding_onsetrate_pcc} are correlation-based correspondences, \ref{fig:vggish_embedding_loudness_l2} and \ref{fig:vggish_embedding_onsetrate_l2} are $L2$-distance based correspondences.}
\label{fig:vggish_embedding_l2}
\end{figure}

\textbf{High-level embeddings vs. onset rate and loudness.}
We explored three high-level activation layers of VGGish model: an embedding layer with 128 neurons and two fully-connected layers with 4096 neurons each. For the embedding layer, we found statistically significant correlations for both onset rate and loudness, and some examples of the corresponding features are shown in Figure \ref{fig:vggish_embedding_l2}. 
In the first fully-connected layer we discovered that neuron \#1964 has an outstanding correlation with loudness (with correlation coefficient $ r = 0.76$). For CNN-AT we found that activation \#259 corresponds to onset rate. 

\textbf{Low-level feature correspondences.}
We found a number of interesting activation maps which look similar to HPSS decomposition in the first convolutional layer of VGGish network. The histograms of similarity metrics with respect to activation maps can be found in supplementary materials.\footnote{Supplementary materials (high resolution figures, code and more examples) are located at \url{https://goo.gl/jM3jZM}. }
The second convolutional layer of VGGish network does not have a strong correspondence to HPSS decomposition even though some linear combinations of activation maps could be similar. 

For CNN-AT network we examine the second convolutional layer and we observe that similarity metric histograms for HPSS decomposition are not consistent which might be related to a higher false matching rate between decompositions and activation maps. 
Finally, the first layers of MM-CNN architecture represent a strongly filtered spectrograms, so we presume that the tall rectangular filters of this architecture are similar to band-pass filters.  

\section{Conclusion}

Even if the models we investigate are complex and allow to construct features in a very different way than traditional methods, the correspondences between hand-crafted features and activations provide insights for better understanding of the internal representations of CNNs. We believe that the proposed methodology can be applied to identify important neurons in other tasks and architectures.

\section{Acknowledgement}

  This work has received funding from the Spanish Ministry of Economy and Competitiveness under the Maria de Maeztu Units of Excellence Programme (MDM-2015-0502) and the European Research Council (ERC)
under the European Union’s Horizon 2020 research and innovation programme (grant 770376, TROMPA). We gratefully acknowledge the support of NVIDIA Corporation with the donation of the Titan X GPU used for this research.

% Section 1. paragraph 2: I think you should provide a short description of (1) and (2). 
%Section 1. Paragraph 4: having dozens of different hand --> a number of well-established mid-level audio features that have been proposed....I would add a reference to a paper, e.g. https://www.nowpublishers.com/article/Details/INR-042
%Section 2. Hand-crafted features. Please clarify that they are somehow complementary and some of them are scalar values while other ones are vectors. How do you compare both in a different way? 
%I would also like to know why you did not use MFCCs or any other timbre feature.  
%Section 2. We consider two similarity metrics --> we consider to SOTA/ well established metrics that are valid to compare different feature sets. 
%Section 3. Do we need to provide 6 decimals in r value? Could you have a small tables including different significant correlations? 

\bibliography{example_paper}
\bibliographystyle{icml2018}

\end{document}